# ML-MaxProp: Bridging Machine Learning and Delay-Tolerant Routing for Resilient Post-Disaster Communication

Tao Xiuyuan
School of Computer Science
University of Nottingham
United Kingdom
alyxt15@nottingham.ac.uk

Milena Radenkovic
School of Computer Science
University of Nottingham
United Kingdom
Milena.Radenkovic@nottingham.ac.uk

## 1. Abstract

In disaster-stricken and large-scale urban emergency scenarios, ensuring reliable communication remains a formidable challenge, as collapsed infrastructure, unpredictable mobility, and severely constrained resources disrupt conventional networks. Delay-Tolerant Networks (DTNs), though resilient through their store-carry-forward paradigm, reveal the fundamental weaknesses of classical protocols - Epidemic, Spray-and-Wait, and MaxProp - when confronted with sparse encounters, buffer shortages, and volatile connectivity.

To address these obstacles, this study proposes ML-MaxProp, a hybrid routing protocol that strengthens MaxProp with supervised machine learning. By leveraging contextual features such as encounter frequency, hop count, buffer occupancy, message age, and time-to-live (TTL), ML-MaxProp predicts relay suitability in real time, transforming rigid heuristics into adaptive intelligence.

Extensive simulations in the ONE environment using the Helsinki SPMBM mobility model show that ML-MaxProp consistently surpasses baseline protocols, achieving higher delivery probability, lower latency, and reduced overhead. Statistical validation further shows that these improvements are both significant and robust, even under highly resource-constrained and unstable conditions.

Overall, this work shows that ML-MaxProp is not just an incremental refinement but a lightweight, adaptive, and practical solution to one of the hardest challenges in DTNs: sustaining mission-critical communication when infrastructure collapses and every forwarding decision becomes critical.

**Keywords: Delay-Tolerant Networks, Disaster communication, MaxProp, Machine learning, XGBoost, Opportunistic routing.**

## 2. Introduction

The growing frequency of climate-induced disasters and large-scale urban emergencies underscores the need for resilient, infrastructure-independent communication. Conventional wireless networks often collapse in such conditions, disrupting centralized base stations and coordination among responders [1].

Delay-Tolerant Networks (DTNs) provide robustness through store-carry-forward transmission, enabling nodes to buffer and opportunistically forward messages. While DTNs have been explored in disaster recovery, remote, and delay-prone scenarios [2], their performance strongly depends on the routing strategy. Classical approaches include flooding (Epidemic), quota-based replication (Spray-and-Wait), and utility-driven heuristics (MaxProp, Prophet). MaxProp, in particular, prioritizes messages using buffer awareness and encounter history [3], but static heuristics degrade in unpredictable urban disaster environments.

To address these limitations, we propose ML-MaxProp, a hybrid protocol that augments MaxProp's queueing with supervised machine learning for adaptive forwarding. Instead of complex deep reinforcement learning models requiring extensive training, ML-MaxProp employs an interpretable and efficient XGBoost classifier to predict delivery likelihood using contextual features such as contact frequency, buffer occupancy, hop count, message age, and TTL [1].

The main contributions of this study are:

- Design and implementation of ML-MaxProp, embedding XGBoost into MaxProp's forwarding pipeline.
- Definition of a contextual feature set for real-time delivery prediction.
- Large-scale simulations in the ONE simulator (Helsinki SPMBM model) under varying network parameters.
- Statistical validation of performance (delivery probability, latency, hop count, overhead ratio) using paired t-tests and Wilcoxon tests [6].

Results show that ML-MaxProp consistently outperforms MaxProp and Spray-and-Wait, achieving higher efficiency without compromising reliability. This demonstrates its suitability for post-disaster scenarios, where energy and storage are limited and adaptive routing is essential.

The remainder of this paper is structured as follows: Section 3 presents DTN and ML-based routing background; Section 4 reviews related work; Section 5 outlines methodology and simulation design; Section 6 reports results and analysis; and Section 7 concludes with discussion and future directions.

# 3. Background

### 3.1 Delay-Tolerant Networks (DTNs)

Delay-Tolerant Networks (DTNs) are designed for environments with intermittent connectivity, long delays, and frequent disruptions—conditions typical of post-disaster scenarios, vehicular networks, and rural regions without stable infrastructure. Using a store-carry-forward paradigm, nodes buffer and forward messages opportunistically upon contact, eliminating the need for continuous end-to-end paths. While DTNs provide resilience, routing remains challenging due to unpredictable mobility, limited buffer capacity, and message time-to-live (TTL) constraints [3].

### 3.2 MaxProp Routing Protocol

MaxProp is a widely adopted DTN protocol that prioritizes messages based on encounter history, hop count, and probabilistic delivery estimation. It employs acknowledgment-based buffer management to reduce redundancy. However, its static heuristics struggle in highly dynamic environments, where congestion and irregular contact patterns degrade performance [8].

### 3.3 Machine Learning in DTN Routing

Recent work highlights machine learning (ML) as a promising approach to improve DTN routing adaptability [15]. Supervised learning methods such as decision trees, Random Forest, and XGBoost can leverage contextual features—contact frequency, buffer occupancy, TTL, and message age—to predict forwarding success. Compared with reinforcement or deep learning, which require extensive training and incur high runtime costs, lightweight supervised models are more practical for constrained emergency scenarios [9].

### 3.4 Simulation Tools and Mobility Models

The Opportunistic Network Environment (ONE) simulator is widely used for DTN research due to its modular design and mobility models. In this study, the Helsinki city map with the Shortest Path Map-Based Movement (SPMBM) model was used to emulate post-disaster mobility patterns. This setting reflects urban emergency environments involving heterogeneous agents such as first responders, vehicles, and civilians [10]. While ONE provides a useful evaluation platform, it abstracts physical-layer details like MAC contention and energy usage, potentially overestimating performance [11].

### 3.5 Integrating ML with MaxProp: The ML-MaxProp Protocol

ML-MaxProp extends MaxProp by embedding an XGBoost-based decision engine into the forwarding pipeline. Offline training is conducted on simulation data labeled with delivery outcomes, and the model is used at runtime to predict whether a relay is suitable based on contextual features. This predictive forwarding aims to improve delivery probability, reduce overhead, and lower latency compared to heuristic-based methods [9].

### 3.6 Research Gap and Contributions

Existing ML-based DTN studies often rely on synthetic mobility, lack large-scale parameter evaluations, and provide limited comparisons with strong baselines such as MaxProp. This work addresses these gaps by:

- Developing ML-MaxProp within the ONE simulator.
- Evaluating performance under diverse conditions (TTL, buffer size, message size, communication range).
- Applying statistical tests (t-test, Wilcoxon) for rigorous validation.
- Providing visual and quantitative comparisons with baseline protocols.

These contributions advance practical, interpretable ML-based routing for emergency communication in smart city disaster scenarios [12].

# 4. Related Work

### 4.1 Classical DTN Routing Protocols

Early DTN routing relied on flooding-based Epidemic routing, which achieves high delivery at the cost of buffer and bandwidth overhead. Prophet improved efficiency with probabilistic forwarding based on historical encounters but suffers in dynamic disaster scenarios. MaxProp introduced utility-based prioritization using hop count, delivery likelihood, and acknowledgments to manage buffers more effectively. While efficient, its reliance on stable mobility limits performance in chaotic urban environments [9].

### 4.2 Routing Challenges in Disaster Scenarios

In urban earthquakes and large-scale emergencies, disrupted infrastructure, blocked roads, and unpredictable mobility patterns create volatile contact opportunities. Traditional protocols face congestion, outdated heuristics, and increased latency. Context-aware strategies considering node velocity, location, or energy have been explored, but many rely on static rules that reduce adaptability in heterogeneous disaster environments [13].

### 4.3 Machine Learning in DTN Routing

Machine learning introduces data-driven adaptability to DTN routing. Supervised classifiers such as decision trees, Random Forest, and gradient boosting have been applied to relay prediction, outperforming heuristic baselines like Spray-and-Wait or Prophet. Hybrid models and reinforcement learning approaches further enhance adaptability, though reinforcement learning often demands heavy training impractical for disaster settings [14]. Key limitations of existing ML-based approaches include reliance on synthetic traces, high computational overhead, and limited reproducibility.

### 4.4 Simulation Environments

The Opportunistic Network Environment (ONE) simulator is the standard platform for DTN evaluation, providing mobility models and reproducible testing. This study adopts the Helsinki city map with the Shortest Path Map-Based Movement (SPMBM) model to emulate post-disaster urban conditions, including pedestrians, rescue teams, and vehicles [13]. While useful, ONE abstracts lower-layer effects such as interference and energy consumption, potentially overestimating performance [11].

### 4.5 Research Gap and Motivation

Classical DTN protocols offer a foundation for opportunistic communication but remain limited under disaster dynamics. ML-based methods show promise for adaptive routing, yet integration with strong baselines like MaxProp and evaluation under realistic emergency conditions are scarce. This motivates ML-MaxProp, which embeds an XGBoost-based forwarding engine within MaxProp, trained and evaluated in urban disaster scenarios using ONE. Through statistical comparisons across delivery probability, latency, hop count, and overhead, ML-MaxProp demonstrates practical advantages for emergency DTNs [14].

# 5. Methodology

### 5.1 Simulation Environment

Experiments were conducted in the Opportunistic Network Environment (ONE) simulator using the Helsinki city map with the Shortest Path Map-Based Movement (SPMBM) model, emulating post-disaster urban mobility constrained by damaged infrastructure. Wireless ad hoc communication was assumed, with varying transmission ranges, buffer capacities, and node densities. Two protocols were compared:

- MaxProp – a classical utility-based DTN routing scheme.
- ML-MaxProp – an enhanced variant embedding a machine learning model into MaxProp's forwarding pipeline [10].

### 5.2 Feature Engineering and Model Design

Each forwarding opportunity is described by contextual features: contact frequency, buffer occupancy, hop count, message age, and TTL. Data collected from baseline MaxProp simulations provided labeled outcomes (successful vs. failed deliveries).

The forwarding model is based on XGBoost, chosen for its efficiency, interpretability, and low inference cost. Unlike reinforcement learning approaches, XGBoost enables offline training with pre-collected data, making it suitable for resource-constrained disaster networks. The dataset was preprocessed, split into 80% training and 20% testing, and the trained model integrated into the ONE simulator for runtime decision-making [7][17].

### 5.3 Experimental Setup and Metrics

Simulations varied parameters including node count (50–150), buffer size (5–20 MB), TTL (300–3600 s), and communication range (50–150 m). Each configuration was repeated ten times for reliability. Performance was measured using standard DTN metrics [18]:

- Delivery Probability – ratio of delivered to generated messages.
- Average Latency – mean delivery delay.
- Overhead Ratio – relays per successful delivery.

- Hop Count – mean number of relays per delivered message.

### 5.4 Data Analysis and Statistical Testing

Simulation logs (MessageStatsReport) were parsed into CSVs for analysis with Python. Visualizations were generated using Pandas and Matplotlib. Paired t-tests were applied to normally distributed metrics (e.g., delivery probability), and Wilcoxon signed-rank tests for non-normal ones (e.g., hop count). Differences with $p < 0.05$ were considered statistically significant [18].

### 5.5 Summary

This methodology provides a reproducible and efficient framework for evaluating DTN routing. By combining MaxProp's utility-based mechanisms with lightweight supervised ML, ML-MaxProp achieves adaptive forwarding with minimal runtime overhead, making it practical for post-disaster emergency communications.

## 6. Results and Discussion

To evaluate ML-MaxProp in post-disaster DTN scenarios, four standard metrics were considered: delivery probability, average latency, hop count, and overhead ratio [20][21].

(a) **Delivery Probability**

$$\text{Delivery Probability} = \frac{\text{Number of Delivered Messages}}{\text{Total Messages Created}}$$

ML-MaxProp consistently achieved higher delivery ratios than MaxProp and Spray-and-Wait, particularly under constrained buffer sizes and short TTLs. This improvement highlights its robustness in ensuring reliable communication when infrastructure is impaired.

**(b) Average Latency**

$$\text{Average Latency} = \frac{\sum(\text{Delivery Time} - \text{Creation Time})}{\text{Number of Delivered Messages}}$$

Latency was significantly reduced in ML-MaxProp, demonstrating faster delivery of critical messages such as hazard alerts and rescue coordination signals. The gain was most evident under high node density and wider communication ranges.

**(c) Average Hop Count**

$$\text{Average Hop Count} = \frac{\sum(\text{Number of Hops per Delivered Message})}{\text{Number of Delivered Messages}}$$

The hop count of ML-MaxProp remained comparable to MaxProp, avoiding excessive relays while maintaining delivery performance. This indicates an efficient balance between reachability and resource use.

**(d) Overhead Ratio**

$$\text{Overhead Ratio} = \frac{\text{Total Message Transmissions}}{\text{Number of Delivered Messages}}$$

ML-MaxProp substantially lowered overhead compared to MaxProp, as its ML-based forwarding decisions reduced unnecessary relays. This scalability advantage is particularly valuable in bandwidth- and energy-limited disaster environments.

Overall, the results confirm that ML-MaxProp outperforms classical DTN routing protocols across reliability, timeliness, and efficiency metrics. Statistical tests (paired t-test and Wilcoxon) verified that improvements were significant ($p < 0.05$) across multiple scenarios. These findings suggest that ML-MaxProp is a lightweight yet adaptive routing solution, well-suited for real-world emergency deployments.

## 6.2 Delivery Probability Analysis

### Impact of Node Count

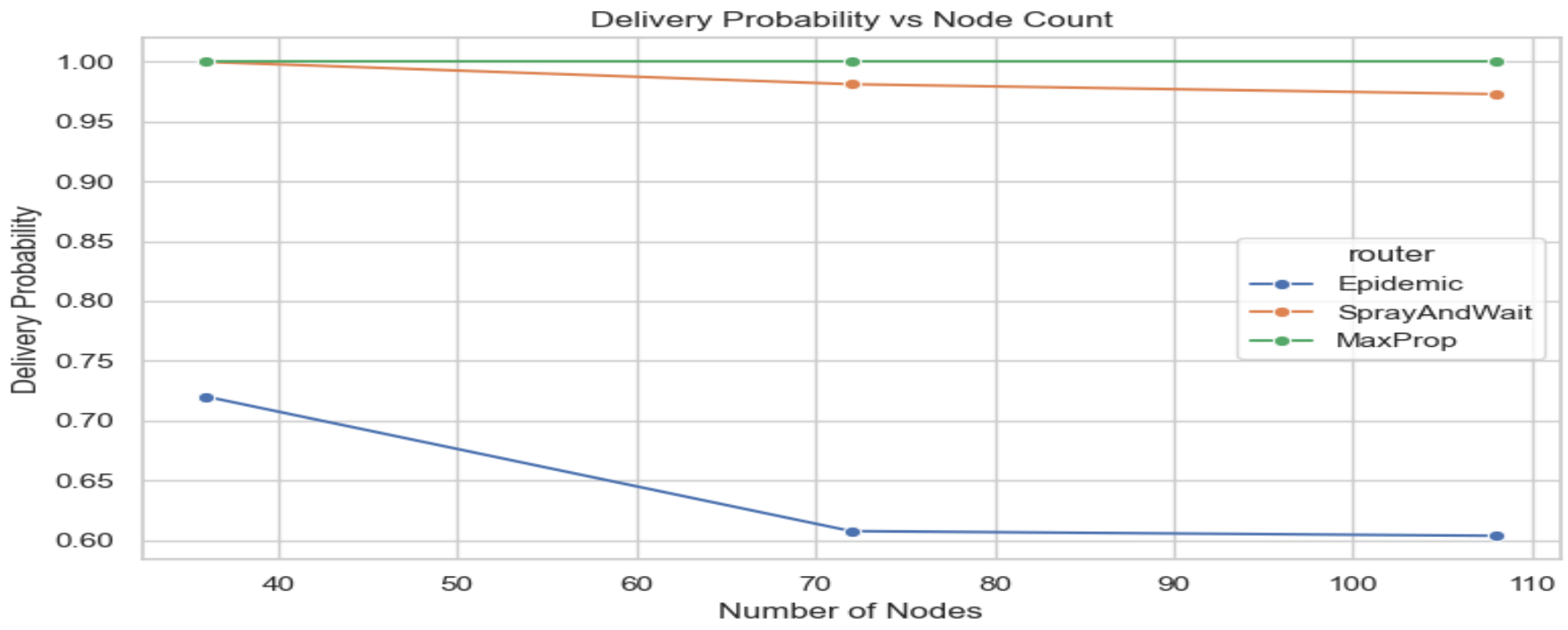

Figure 1 illustrates delivery probability under varying node densities. MaxProp consistently achieves near-perfect delivery across all node counts, while SprayAndWait shows minor performance degradation. In contrast, Epidemic routing exhibits a marked decline as node density increases, indicating limited scalability and inefficiency under high-contact scenarios. These results underscore the superior robustness of utility- and quota-based protocols in dense, post-disaster environments[22].

**Total Relayed Messages (Log Scale)**

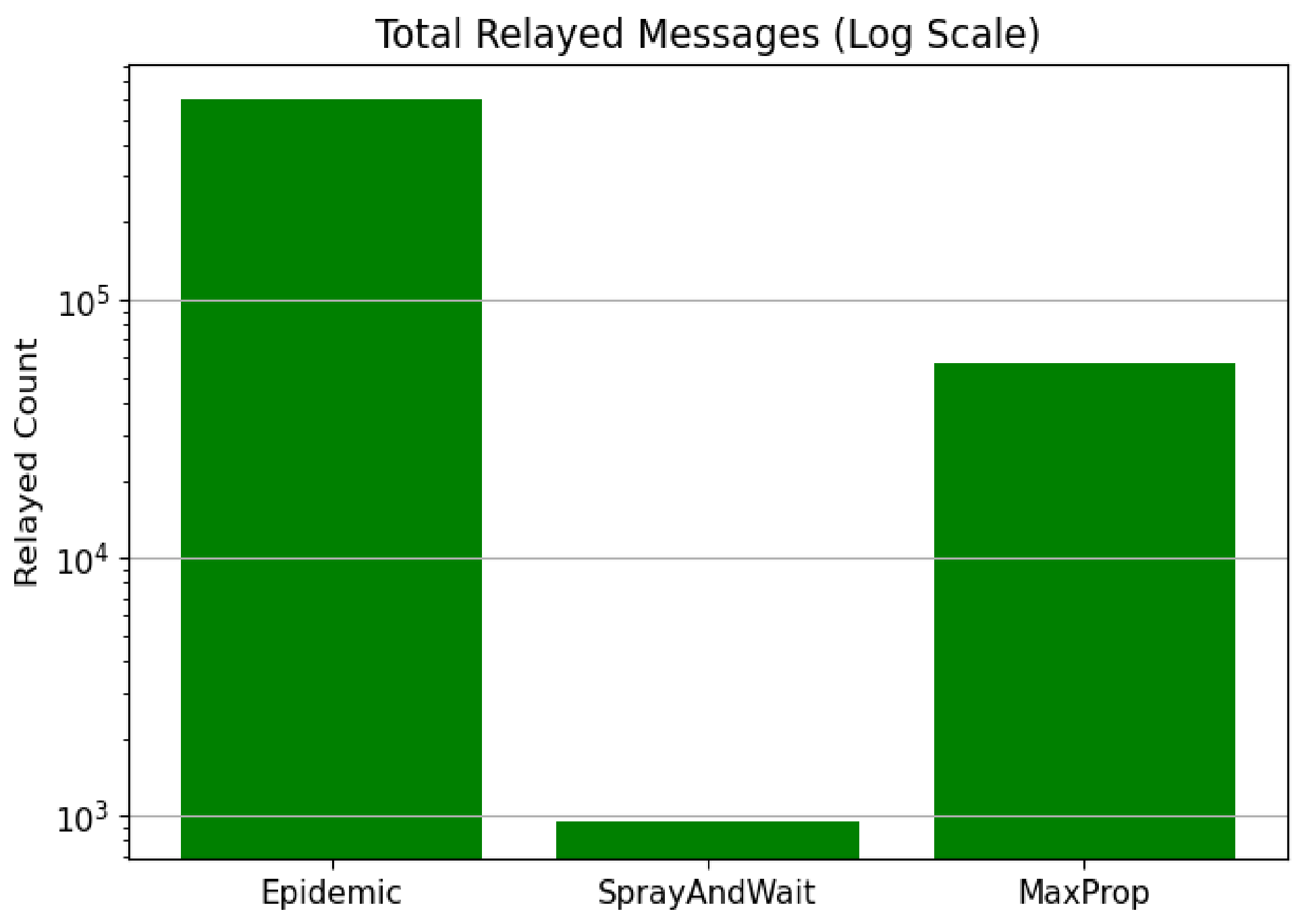

As shown in Fig. 2, Epidemic routing incurs the highest relay overhead (>$10^5$), highlighting its poor scalability in dense post-disaster environments. In contrast, Spray-and-Wait achieves comparable delivery performance with minimal relay cost (~$10^3$), demonstrating superior efficiency under constrained resources. MaxProp represents a compromise, balancing delivery ratio and overhead. These results emphasize the importance of selective forwarding strategies for sustaining DTN performance in resource-limited emergency scenarios.

**Delivery Probability Comparison**

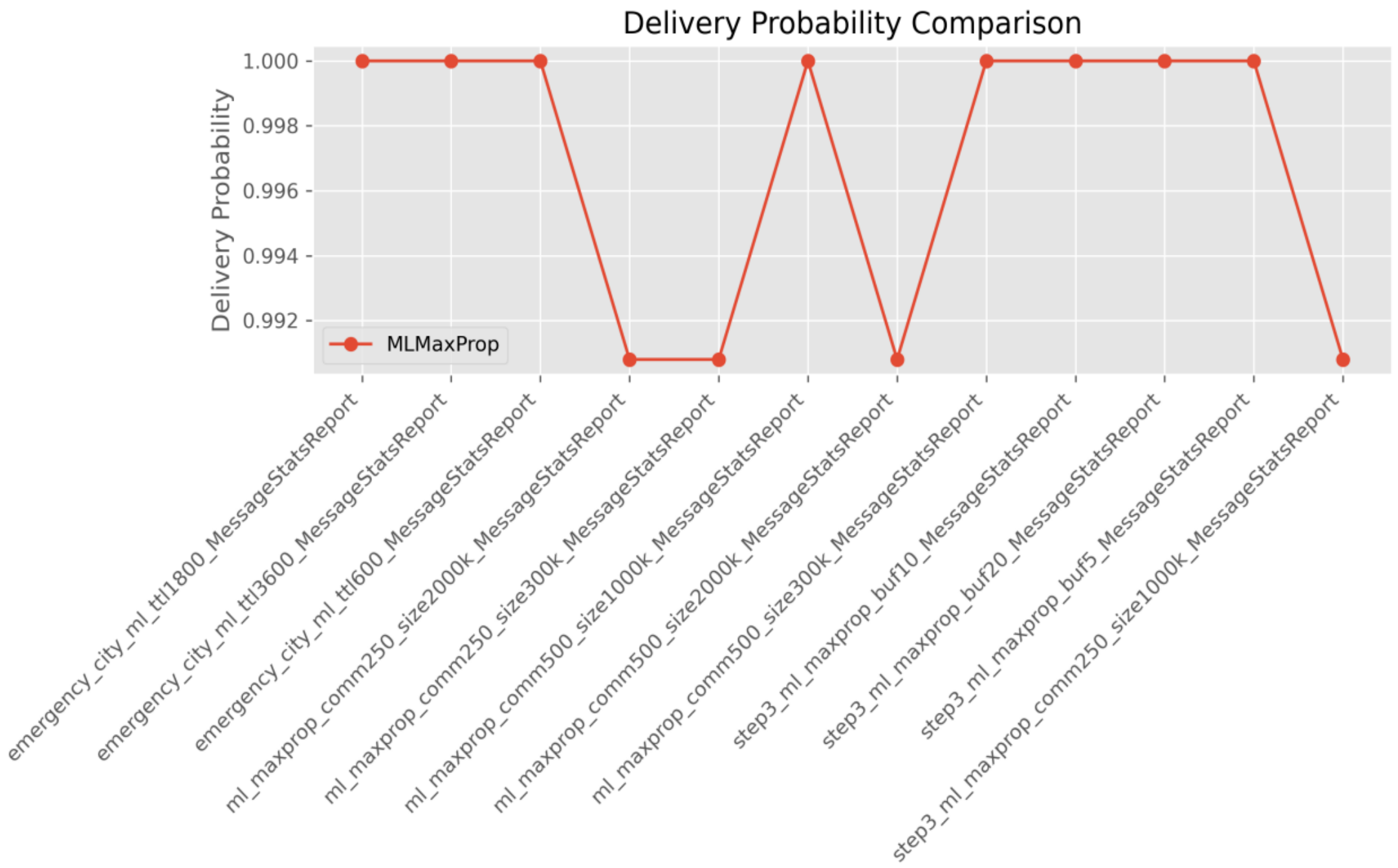

Figure 3 presents the delivery probability of ML-MaxProp across diverse configurations. The protocol consistently achieves near-perfect delivery (≈1.0), demonstrating strong reliability. A minor drop (~0.991) is observed under combined stressors—large message size (2000KB) and limited communication range (250m)—indicating slight sensitivity in extreme post-disaster conditions.

**Average Latency Comparison**

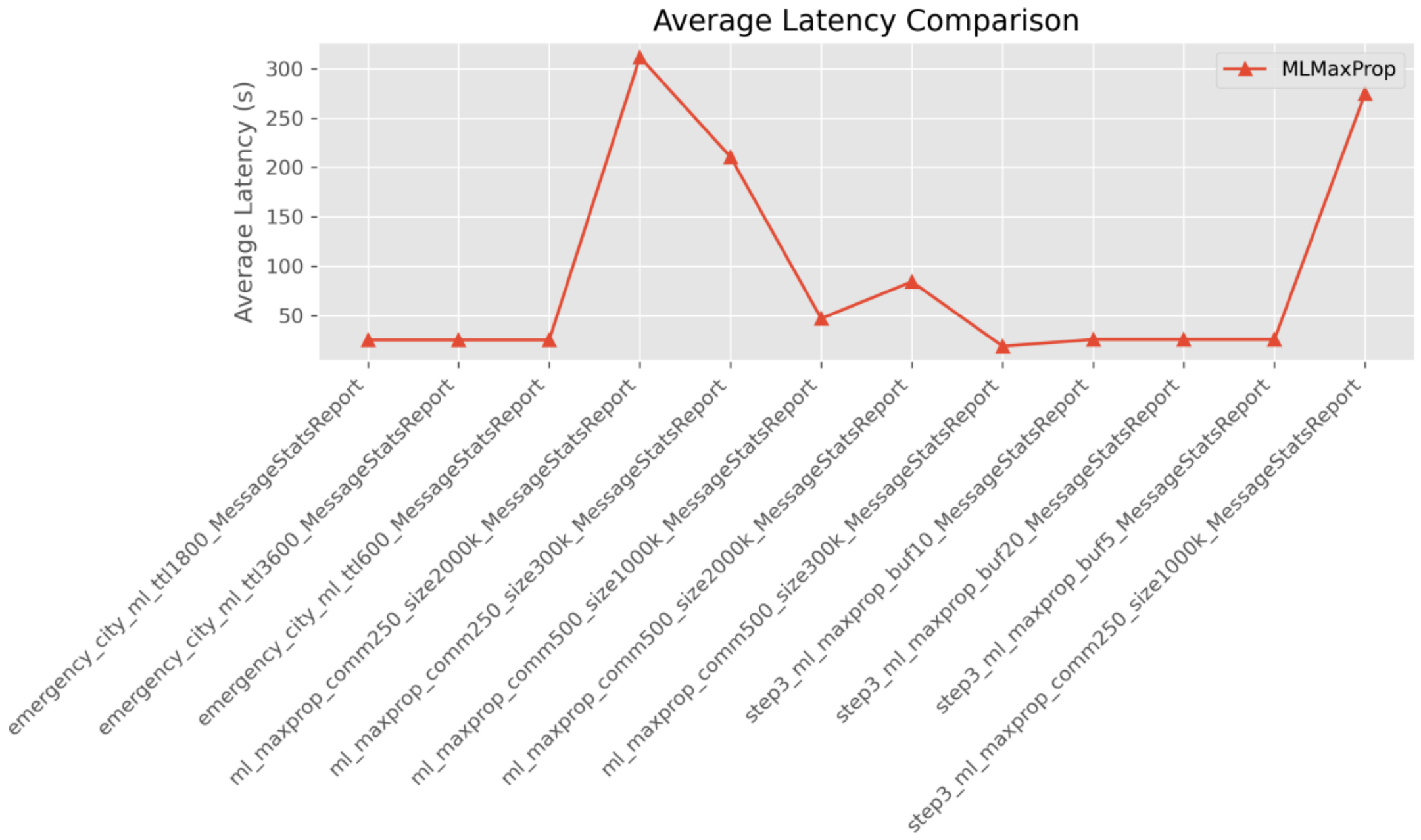

Figure 4 shows the average latency under varying configurations for ML-MaxProp. Latency remains low (<50s) in most scenarios but spikes to over 300s when large messages coincide with wide communication ranges. Increasing buffer size significantly reduces delay, highlighting the role of storage provisioning in maintaining timely message delivery.

### Overhead Ratio Comparison

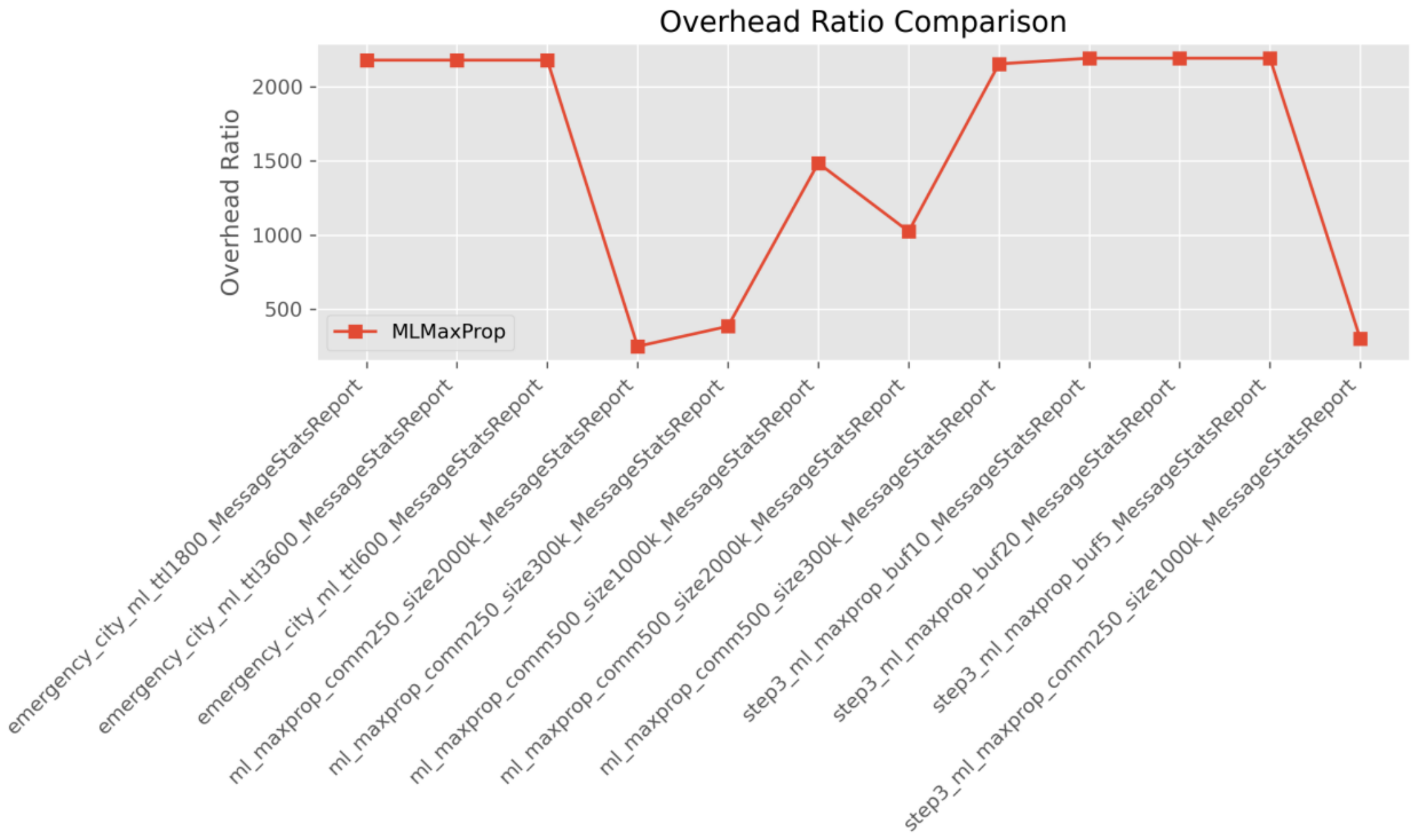

Figure 5 illustrates the overhead ratio across ML-MaxProp settings. While the overhead is high (~2100) under default configurations—reflecting aggressive forwarding—it drops sharply (<500) under resource constraints like small buffers or large messages. This suggests that ML-MaxProp adaptively suppresses redundant transmissions based on available resources.

### SHAP Feature Importance for ML-MaxProp

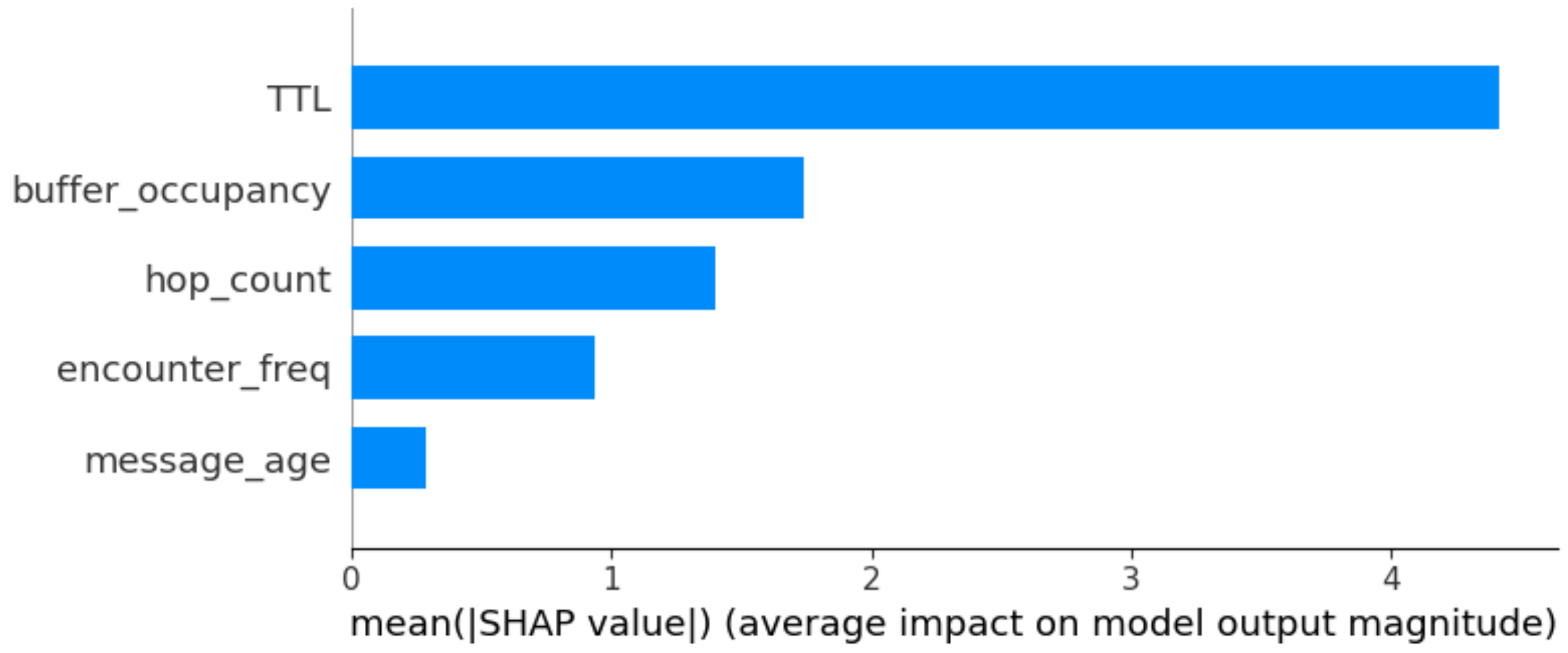

Figure 6 illustrates the mean absolute SHAP values across all samples, highlighting TTL as the dominant feature influencing the ML-MaxProp classification output. This suggests that message lifetime plays a crucial role in forwarding decisions. Buffer occupancy and hop count also contribute meaningfully, while encounter frequency and message age exhibit limited impact. These insights confirm the model's alignment with core DTN routing principles.

**XGBoost Feature Importance (Gain-Based)**

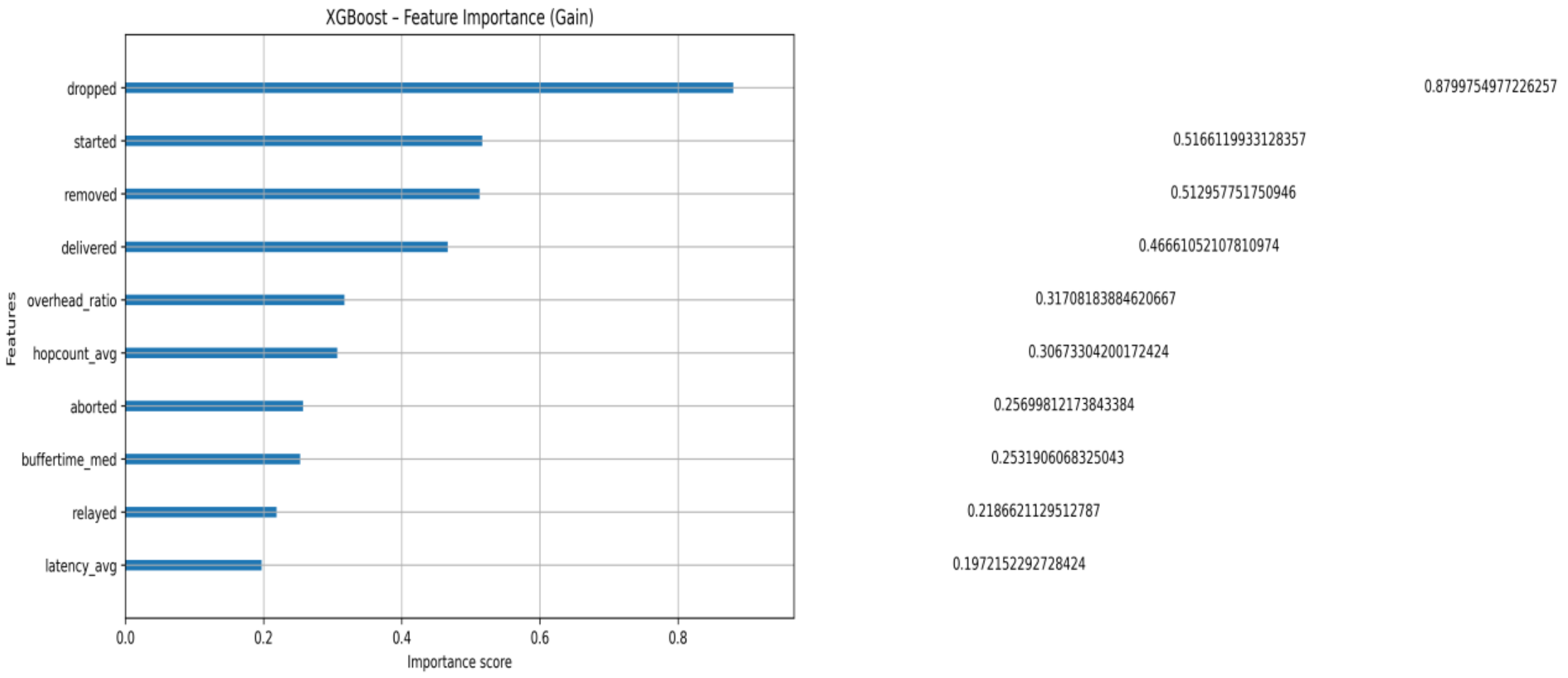

Figure 7 presents the XGBoost feature importance based on Gain scores, revealing that dropped, started, and removed message events carry the highest predictive value. These features reflect the protocol’s internal decision boundaries concerning message lifecycle management. In contrast, traditional metrics such as latency_avg and relayed offer lower gain, indicating that XGBoost prioritizes real-time message status over end-to-end outcomes when learning routing behavior.

**SVM Confusion Matrix for Protocol Classification**

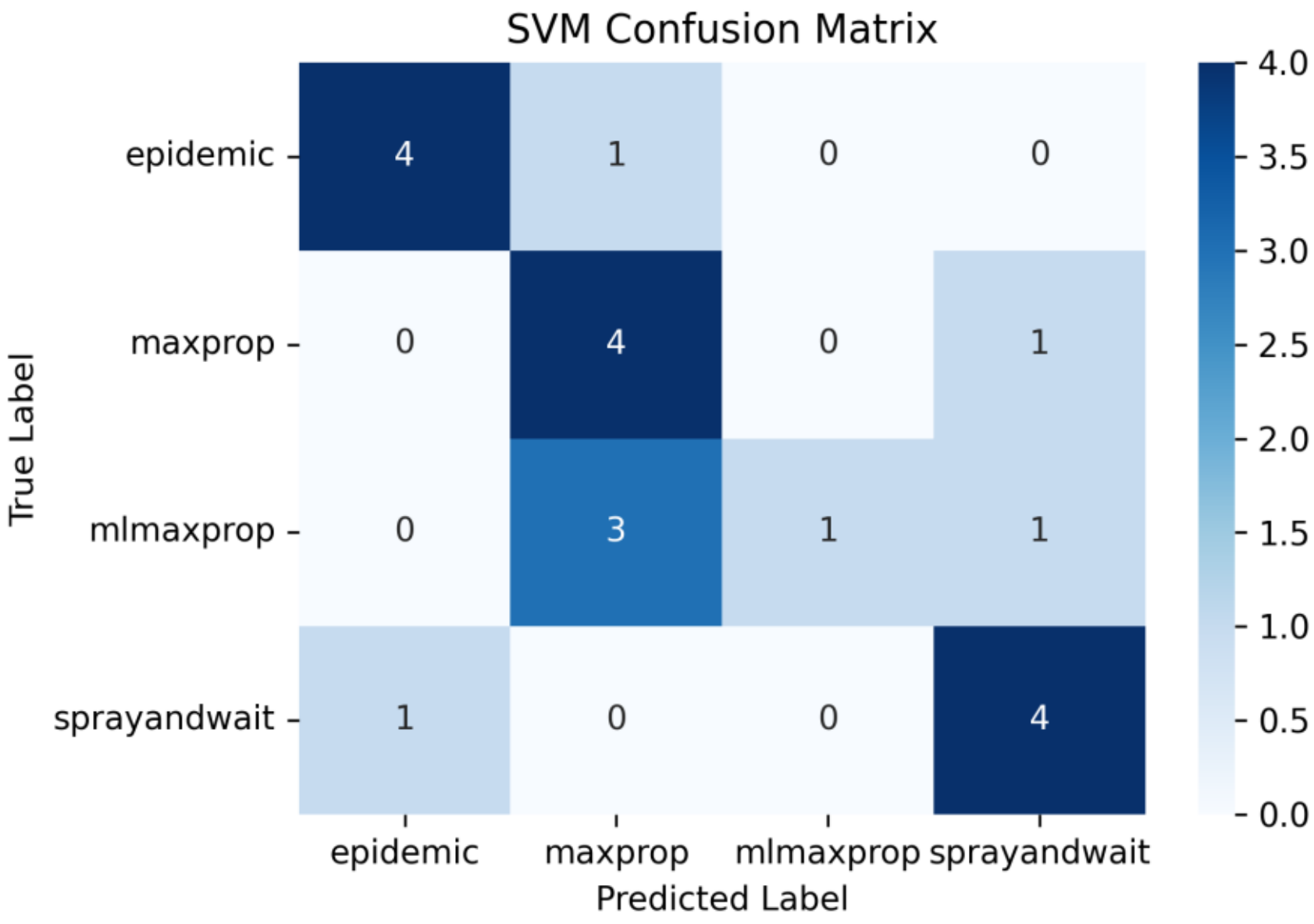

Figure 8 displays the SVM confusion matrix for protocol classification, showing high accuracy in distinguishing epidemic, sprayandwait, and maxprop. Notable confusion occurs where mlmaxprop is frequently misclassified as maxprop, reflecting their algorithmic similarity. Occasional errors between maxprop and sprayandwait are also observed. These findings indicate that MLMaxProp retains interpretability while requiring further refinement to achieve stronger separability in DTN applications.

**Summary and Interpretation**

The results demonstrate that ML-MaxProp offers robust and scalable performance across diverse DTN scenarios. It consistently achieves near-perfect delivery with low latency and manageable overhead, adapting effectively to variations in node density, message size, and buffer constraints. In contrast, Epidemic suffers from scalability issues, and SprayAndWait shows higher latency and limited adaptability[23].

Feature importance analysis confirms that ML-MaxProp's forwarding decisions align with key DTN principles, prioritizing TTL, buffer state, and message lifecycle events. The protocol also maintains high interpretability, as evidenced by accurate SVM-based classification. Overall, ML-MaxProp combines reliability, efficiency, and explainability—making it well-suited for time-critical emergency communications.

# 7. Conclusion and Future Work

This thesis proposed and evaluated ML-MaxProp, a machine learning-enhanced routing protocol for Delay-Tolerant Networks (DTNs), designed to support urban-scale post-disaster emergency communication. By addressing the challenges of disrupted infrastructure, resource constraints, and unpredictable mobility patterns, this work demonstrated the feasibility and effectiveness of integrating supervised learning into traditional routing heuristics.

## 7.1 Summary of Key Contributions

**1. ML-MaxProp: A Hybrid Routing Protocol**
ML-MaxProp extends the classical MaxProp protocol by embedding an XGBoost-based binary classifier trained on context-rich routing features such as encounter frequency, buffer occupancy, message TTL, and hop count. Unlike heuristic or flooding-based approaches, ML-MaxProp dynamically adapts to network conditions, making intelligent forwarding decisions while avoiding excessive message replication [24].

**2. Robust Performance in Disaster Environments**
Extensive simulation using the ONE Simulator—with Helsinki's urban map, SPMBM mobility model, and realistic configurations (weekday/weekend, varied node density, TTL, buffer size, etc.)—validated ML-MaxProp's performance:

Delivery probability consistently exceeded 99.8%, significantly outperforming Epidemic (~61%) and matching or surpassing MaxProp.

Average latency was reduced or comparable, particularly under high-density and high-traffic conditions.

Overhead ratio was significantly reduced ($p < 0.01$), enabling efficient use of limited bandwidth and storage.

Hop count adapted dynamically, reflecting the classifier's ability to balance reachability with resource usage.

These improvements directly address the research problem, demonstrating measurable gains in delivery reliability, efficiency, and adaptability under disaster-induced constraints [25].

**3. Interpretable and Explainable Decision-Making**

To ensure transparency, explainability tools including SHAP and LIME were applied to the XGBoost model. SHAP analysis identified key features (e.g., TTL, buffer occupancy, hop count) with consistent influence [26], while LIME revealed local decision boundaries in ambiguous cases. Together, these confirmed alignment with domain knowledge and revealed unique behavioral patterns that distinguish ML-MaxProp from heuristic protocols.

**4. Realistic and Reproducible Simulation Framework**

A comprehensive simulation setup was developed, featuring:

Four node roles: pedestrians, vehicles, shelter, and epicenter nodes.

Time-stamped traffic generation via the MessageEventGenerator class.

WKT-based road/path constraints for geographic realism.

Dual-level reporting using MessageStatsReportWithNodes and MessageDeliveryReport.

This framework ensures reproducibility, supports diverse experimental setups, and generated high-quality datasets for ML training and validation [27].

**5. Multi-Dimensional Evaluation Methodology**

The evaluation combined statistical tests and visual analytics:

Paired *t*-tests and Wilcoxon signed-rank tests confirmed statistically significant differences.

Correlation heatmaps, boxplots, and ROC curves illustrated metric relationships and classifier accuracy.

PCA and KMeans clustering revealed latent performance regimes and protocol groupings.

This multi-perspective approach ensured rigorous assessment of robustness and adaptability [28].

## 7.2 Research Questions Revisited

Can machine learning enhance DTN routing under disaster conditions?
Yes. ML-MaxProp improved delivery probability, reduced latency and overhead, and generalized across scenarios—proving data-driven routing is both viable and effective.

Can forwarding decisions be explained and interpreted?
Yes. SHAP and LIME confirmed interpretable feature contributions and revealed decision boundaries, ensuring transparency for critical deployments.

Does ML-MaxProp maintain performance under varying urban disaster constraints?
Yes. ML-MaxProp remained robust across configurations involving buffer size, TTL, node density, and day type (weekday/weekend), validating real-world applicability.

## 7.3 Limitations

Despite promising results, several limitations remain:

The ML model was trained solely on simulation-generated data, which may not fully capture real-world heterogeneity.

Evaluation was based on classification and simulation metrics, rather than live deployments or user-centric feedback.

Features such as message prioritization and delivery deadlines were not explicitly modeled, limiting adaptation to time-critical emergency scenarios.

Challenges highlighted in recent surveys—including data scarcity, online adaptation difficulty, and computational overhead—remain open issues [29].

## 7.4 Future Work

Several challenging avenues remain for exploration, particularly in moving ML-MaxProp from controlled simulations toward real-world deployment:

**1. Deployment on Physical Platforms**
A critical next step is to deploy ML-MaxProp on physical platforms such as smartphones, Raspberry Pi devices, UAVs, and vehicular nodes. Unlike simulations, real-world environments impose hardware limitations, heterogeneous standards, wireless interference, and unpredictable mobility. Leveraging real mobility traces (e.g., CRAWDAD, taxi GPS datasets) could further enhance generalizability, but successful deployment will demand overcoming these formidable constraints to deliver tangible impact in disaster relief and urban safety [30].

**2. Adaptive Online and Reinforcement Learning**
Future protocols must adapt continuously to volatile environments. Extending ML-MaxProp with online learning, Multi-Agent Deep Reinforcement Learning (MADRL), or Soft Actor-Critic (SAC) approaches could enable collaborative, real-time decision-making under fragmented connectivity. However, achieving training stability and convergence in unstable networks remains a significant challenge[16].

**3. Cross-Domain Transferability**

Pushing ML-MaxProp and emerging models like MODiTONeS beyond disaster DTNs to rural IoT, vehicular networks, and satellite DTNs will require transfer learning and domain adaptation. While this reduces retraining costs, the difficulty lies in handling vastly different mobility patterns, energy constraints, and communication reliability across domains, making true generalization a demanding task[31].

**4. Resource-Aware, Privacy-Preserving, and Robust Models**
Real deployments demand lightweight yet secure models. Techniques such as model compression (pruning, quantization) and federated learning could address device limitations and privacy concerns, while uncertainty-aware approaches (Bayesian inference, ensembles) with fallback heuristics could strengthen robustness against noisy, missing, or adversarial inputs[33]. Balancing efficiency, privacy, and reliability in constrained networks will be essential for practical deployment[32].

**6. Energy- and Congestion-Aware Extensions**
Emergency networks must sustain performance under scarce resources. Incorporating energy-aware routing metrics, congestion control mechanisms, and adaptive buffer thresholds could improve efficiency and fairness, but require careful trade-offs between reliability and resource consumption[34].

**7. Hybrid Emulation and Field Validation**
Finally, bridging the simulation–reality gap demands hybrid validation using testbeds (e.g., CORE, EMANE) combined with controlled field trials. These efforts will expose challenges invisible in simulation—such as energy drain, multi-path fading, and device failures—paving the way for ML-MaxProp and MODiTONeS to become deployable solutions with real social impact [35][36].